\documentclass[aps,prl,twocolumn,groupedaddress]{revtex4-1}
\usepackage{epsfig}
\usepackage{graphics}
\usepackage{graphicx}
\usepackage{lineno} 
\usepackage{makecell}

\begin{document}

%\preprint{}

\title{ Low-$p_T$ $e^{+}e^{-}$ pair production in Au$+$Au collisions at $\sqrt{s_{NN}}$ = 200 GeV and U$+$U collisions at $\sqrt{s_{NN}}$ = 193 GeV at STAR}


%\author{STAR Collaboration}
%\thanks{}
%\affiliation{}

\author{
J.~Adam$^{9}$,
L.~Adamczyk$^{1}$,
J.~R.~Adams$^{31}$,
J.~K.~Adkins$^{21}$,
G.~Agakishiev$^{19}$,
M.~M.~Aggarwal$^{33}$,
Z.~Ahammed$^{56}$,
N.~N.~Ajitanand$^{44}$,
I.~Alekseev$^{17,28}$,
D.~M.~Anderson$^{46}$
% ...
}

\address{$^{1}$AGH University of Science and Technology, FPACS, Cracow 30-059, Poland}
% ...
\address{$^{9}$Creighton University, Omaha, Nebraska 68178}
% ...
\address{$^{17}$Alikhanov Institute for Theoretical and Experimental Physics, Moscow 117218, Russia}
% ...
\address{$^{19}$Joint Institute for Nuclear Research, Dubna, 141 980, Russia}
% ...
\address{$^{21}$University of Kentucky, Lexington, Kentucky 40506-0055}
% ...
\address{$^{28}$National Research Nuclear University MEPhI, Moscow 115409, Russia}
% ...
\address{$^{31}$Ohio State University, Columbus, Ohio 43210}
% ...
\address{$^{33}$Panjab University, Chandigarh 160014, India}
% ...
\address{$^{44}$State University of New York, Stony Brook, New York 11794}
% ...
\address{$^{46}$Texas A\&M University, College Station, Texas 77843}
% ...
\address{$^{56}$Variable Energy Cyclotron Centre, Kolkata 700064, India}
% ...


\collaboration{STAR Collaboration}

\date{\today}

%\linenumbers

\begin{abstract}

We report first measurements of $e^{+}e^{-}$ pair production in the mass region 0.4 $<M_{ee}<$ 2.6 GeV/$c^{2}$ at low transverse momentum ($p_T<$ 0.15 GeV/$c$) in non-central Au$+$Au collisions at $\sqrt{s_{NN}}$ = 200 GeV and U$+$U collisions at $\sqrt{s_{NN}}$ = 193 GeV. Significant enhancement factors, expressed as ratios of data over known hadronic contributions, are observed in the 40-80\% centrality of these collisions. The excess yields peak distinctly at low-$p_T$ with a width ($\sqrt{\langle p^2_T\rangle}$) between 40 to 60 MeV/$c$. The absolute cross section of the excess depends weakly on centrality while those from a theoretical model calculation incorporating an in-medium broadened $\rho$ spectral function and radiation from a Quark Gluon Plasma or hadronic cocktail contributions increase dramatically with increasing number of participant nucleons. Model calculations of photon-photon interactions generated by the initial projectile and target nuclei describe the observed excess yields but fail to reproduce the $p^{2}_{T}$ distributions. 
\end{abstract}

%\keywords{}

\maketitle


%\section{Introduction}
%\subsection{}
%\subsubsection{}

%% Introduction
A major goal of the Relativistic Heavy-Ion Collider (RHIC) is to study properties of the deconfined state of partonic matter, known as the Quark Gluon Plasma (QGP)~\cite{qgpRHIC, qgpTheory}. Dileptons play a crucial role in studying such matter because they are produced during the entire evolution of the hot, dense medium while not being subject to strong interactions with it. Previous dilepton measurements over a wide $p_T$ region at the Super Proton Synchrotron (SPS)~\cite{ceresDilepton, na60Dilepton} and RHIC~\cite{starDilepton200, starDilepton19.6, phenixDilepton} showed a significant enhancement with respect to known hadronic sources in the mass region below $\sim$0.7 GeV/$c^{2}$. The observed excess can be consistently described by model calculations that incorporate an in-medium broadening of the $\rho$ spectral function~\cite{broadenRhoTheory}. 

Strong electromagnetic fields arising from the relativistic contraction and large amount of charges in the nuclei generate a large flux of high-energy quasi-real photons~\cite{photonInteractionTheory, starlightGenerator}. Dileptons can also be produced via these photon interactions~\cite{photonInteractionTheory}, such as photon-photon and photonuclear processes. In the photon-photon process, virtual photons emitted from the by-passing nuclei interact to generate dileptons ($\gamma\gamma\rightarrow\ell^+\ell^-$). In the photonuclear process, virtual photons emitted by one nucleus can interact either with the other whole nucleus (coherent process) or with individual nucleons in the other nucleus (incoherent process) to produce vector mesons ($\gamma+A \rightarrow V+A$), which then decay into dileptons~\cite{VMDmodel}. Dilepton production from either photon-photon or coherent photonuclear processes are known to be distinctly peaked at very low transverse momenta ($p_T$)~\cite{photonInteractionTheory}. The photon interaction processes have been extensively studied in ultra-peripheral collisions (UPCs) with impact parameters larger than twice the nuclear radius~\cite{starUpcRho0, starUpceepair, starUpcRho0t, phenixUpceeandJpsi, aliceUpcJpsi, aliceUpceeandJpsi, cmsUpcJpsi}. The ALICE collaboration recently reported a significant J/$\psi$ excess yield at very low $p_T$ ($p_T<$ 0.3 GeV/$c$) in peripheral Pb$+$Pb collisions at forward rapidity~\cite{aliceLowPtJpsi}, qualitatively explained by coherent photonuclear production mechanisms~\cite{aliceLowPtJpsi, wangmeiTheory}. That explanation implies the existence of an energetic, high-density photon flux produced during the collision from which photon-photon interactions would also occur and contribute to $e^{+}e^{-}$ pair production~\cite{photonInteractionTheory, aliceUpceeandJpsi}. Measurements of $e^+e^-$ pair production at very low $p_T$ from different collision systems and energies become necessary to verify and constrain the photon interactions in heavy-ion collisions with hadronic overlap. In such collisions, the photon-photon interactions could be further used to probe the possible existence of strong magnetic fields trapped in a conducting QGP medium~\cite{cmeConductivity}.

In this Letter, we report centrality and invariant mass dependences of inclusive $e^{+}e^{-}$ pair production at $p_T<$ 0.15 GeV/$c$ in Au$+$Au collisions at $\sqrt{s_{NN}}$ = 200 GeV and U$+$U collisions at $\sqrt{s_{NN}}$ = 193 GeV. The observed excess $e^{+}e^{-}$ yields with respect to the known hadronic sources are presented as a function of centrality and $p_T^2$. Model calculations that include an in-medium modified $\rho$ spectral function and QGP radiation, photon-photon processes, and coherent photonuclear interactions are compared with the measurements.

%% Data and Analysis
The Au$+$Au data used for this analysis were collected by the STAR collaboration~\cite{starExp} during the 2010 and 2011 RHIC runs, while the U$+$U data were collected in 2012. A total of 7.2 $\times$ $10^{8}$ Au$+$Au and 2.7 $\times$ $10^{8}$ U$+$U minimum-bias (0-80\%) events are used. The minimum-bias trigger is defined as a coincidence signal between the east and west vertex position detectors (VPD)~\cite{vpdDet} located at forward pseudorapidities ($\eta$), 4.24 $\le|\eta|\le$ 5.1. The collision centrality is determined by matching the measured charged particle multiplicity within $|\eta|<$ 0.5 with a Monte Carlo Glauber simulation~\cite{cenDef}. The collision vertex is required to be within 30 cm from the STAR detector center along the beam line to ensure uniform detector acceptance, and within 2 cm radius in the plane perpendicular to the beam line. To reject pileup events, the distance between the collision vertex and the vertex reconstructed by the VPD is required to be less than 3 cm along the beam direction.

The main subsystems used for electron (both $e^{+}$ and $e^{-}$) identification are the Time Projection Chamber (TPC)~\cite{tpcDet} and the Time-of-Flight (TOF)~\cite{tofDet} detectors. Tracks reconstructed in the TPC are required to have at least 20 space points (out of a maximum of 45) to ensure sufficient momentum resolution, contain no fewer than 15 space points for the ionization energy loss ($dE/dx$) determination to ensure good $dE/dx$ resolution, and to be matched to a TOF space point. Furthermore, tracks are selected to originate from the collision vertex by requiring the distances of closest approach to this vertex be less than 1 cm. With the combined measurements of $dE/dx$ by the TPC and velocity ($\beta$) by the TOF~\cite{starDilepton200}, a high purity electron sample is obtained. The electron purity for $p_T^{e}>$ 0.2 GeV/$c$ is about 95\% in both Au$+$Au and U$+$U data samples.

The unlike-sign pair distribution (signal and background) at midrapidity ($|y^{ee}|<$ 1), is generated by combining electron and positron candidates with $p_T^{e}>$ 0.2 GeV/$c$ and $|\eta^{e}|<$ 1 from the same event. The background is estimated by combining the same charge sign electrons (like-sign pairs) in the same event. Due to dead areas of the detector and the different bending directions of positively and negatively charged particle tracks in the transverse plane, the unlike-sign and like-sign pair acceptances are not identical. A mixed-event technique is used to correct for the acceptance difference as a function of pair invariant mass ($M_{ee}$) and $p_T$. The raw signal, obtained by subtracting the background from the unlike-sign distribution, is corrected for the detector inefficiency. 

The efficiency is factorized into TPC tracking, matching with TOF, and particle identification as described in detail elsewhere~\cite{starDilepton200}. The TPC tracking efficiency is evaluated via a well-established STAR embedding technique~\cite{embRef}. Simulated electrons, passed through the STAR detector GEANT3 model~\cite{geant3} and detector response algorithms, are embedded into raw minimum-bias triggered events. The efficiency is determined by the rate at which the simulated electrons are found when the events are processed using the standard STAR reconstruction procedure. The TOF matching and particle identification efficiencies are evaluated using a pure electron sample, as described in Ref.~\cite{starDilepton200}. Finally, the electron pair efficiency is determined by convoluting the single electron efficiency as a function of $p_T^{e}$, $\eta^{e}$, and $\phi^{e}$ with the decay kinematics. For the measurements in Au+Au collisions, the efficiency-corrected spectra are obtained separately for 2010 and 2011 data sets, and then combined using the respective statistical errors as weights.

%% Systematic uncertainty
The systematic uncertainties for the raw $e^{+}e^{-}$ signal extraction include: (a) the uncertainty in correcting the acceptance difference between unlike-sign and like-sign distributions, which is 1-8\% depending on the pair $p_T$ and mass; (b) hadron contamination in the electron sample resulting in an uncertainty of less than 4\%. The uncertainties on the detector efficiency correction are 13\%~\cite{starDilepton200} and 10\% for Au$+$Au and U$+$U measurements, respectively. The total systematic uncertainty is determined via the quadratic sum of each component.

%% Cocktail simulation
A Monte Carlo (MC) simulation is performed to account for the contributions from known hadronic sources at late freeze out, also referred to as the hadronic cocktail. The simulation includes the $e^{+}e^{-}$ pair contributions from direct or Dalitz decays of $\pi^{0},~\eta,~\eta{'},~\omega,~\phi,~J/\psi,~\psi{'},~c\bar{c},~b\bar{b}$, and Drell-Yan production. In Au$+$Au collisions, the cocktail components are the same as those in Ref.~\cite{starDilepton200} except for $\eta$, while the $\eta$ component is the same as that in Ref.~\cite{starDirectPhoton}.
In Au+Au collisions, the input cross sections of hadronic cocktail components agree with the measured experimental data~\cite{starDilepton200, phenixEta}. So far, there are no existing measurements of light hadron spectra in U$+$U collisions at $\sqrt{s_{NN}}$ = 193 GeV. However, given that the energy density reached in U$+$U collisions at $\sqrt{s_{NN}}$ = 193 GeV is only about 20\% higher than that in Au$+$Au collisions at $\sqrt{s_{NN}}$ = 200 GeV~\cite{uuEnergyDensity}, the same Tsallis Blast-Wave (TBW) parametrized $p_T$ spectra used in Au$+$Au collisions~\cite{starDilepton200, zeboTBW} are used as inputs to the U$+$U cocktail simulations. The meson yields ($dN/dy$) in U$+$U collisions are derived from those in Au$+$Au collisions. Specifically, the $\pi^{0}$ yield [($\pi^{+}$+$\pi^{-}$)/2] in Au$+$Au collisions~\cite{embRef} scaled by half of the number of participating nucleons ($N_{\rm part}$/2) as a function of $N_{\rm part}$, is fitted with a linear function. The $\pi^{0}$ yields in U$+$U collisions are then determined by this function at given $N_{\rm part}$ values for various centrality bins. For other mesons (except $J/\psi$ and $\psi'$), the ratios of their yields to the $\pi^{0}$ in U$+$U collisions are taken to be the same as that in minimum-bias Au$+$Au collisions, while the $J/\psi$ and $\psi'$ yields per number of binary nucleon-nucleon collisions are assumed to be the same for U+U and Au+Au collisions. The systematic uncertainties on the cocktail are dominated by the experimental uncertainties on the measured particle yields and spectra. Due to lack of measurements, the $p_T$ spectra of the cocktail inputs for $p_T<$ 0.15 GeV/$c$ rely on the TBW extrapolation.

%% Results

\begin{figure}[htbp]
%\begin{center}
\includegraphics[keepaspectratio,width=0.5\textwidth]{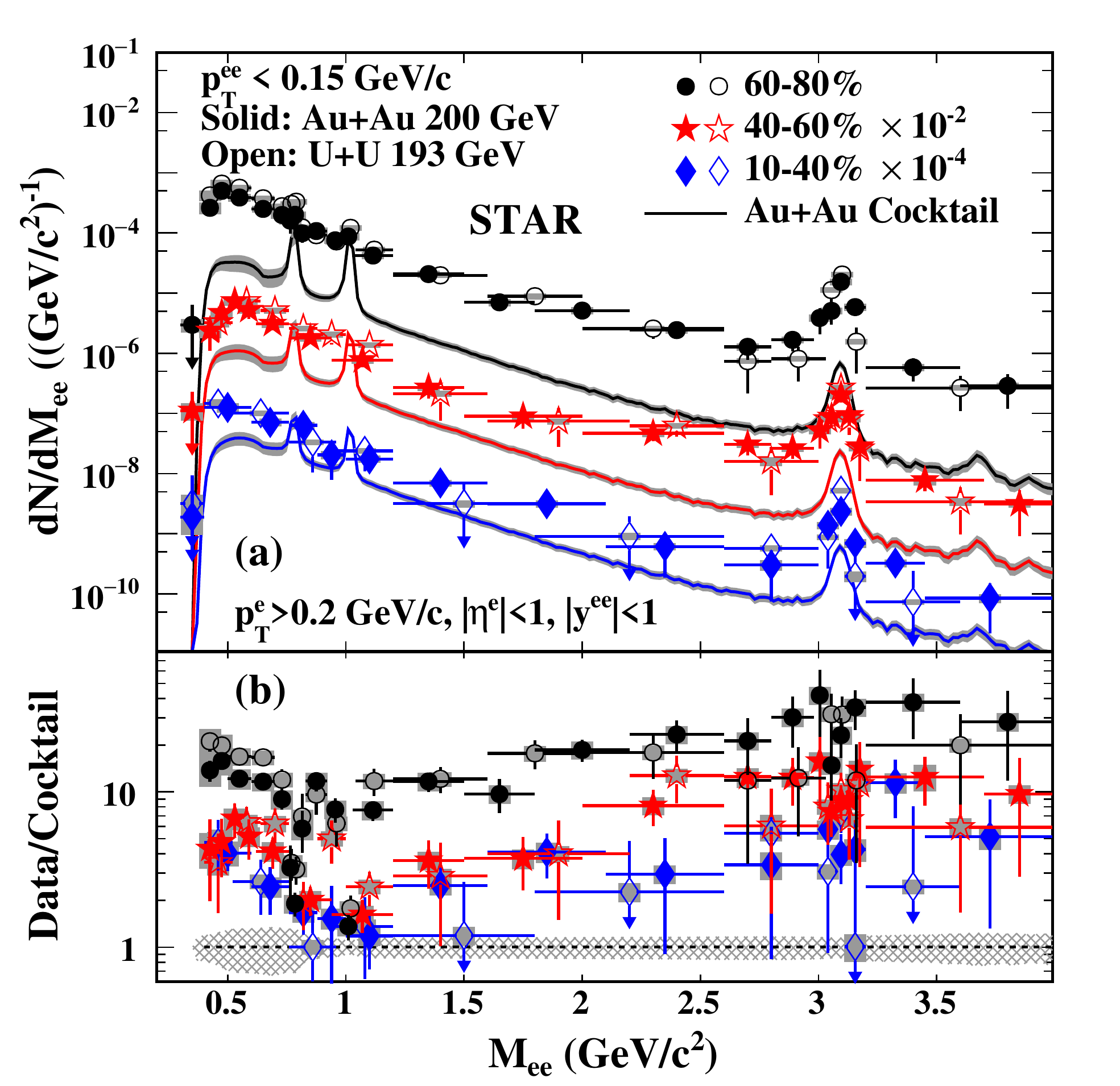}
%\includegraphics[keepaspectratio,angle=270,width=0.5\textwidth]{lowPt_dielectron_massSpectra.pdf}
%\vspace*{-3mm}
\caption{(a) The centrality dependence of $e^{+}e^{-}$ invariant mass spectra within the STAR acceptance from Au$+$Au collisions and U$+$U collisions for pair $p_T<$ 0.15 GeV/$c$. The vertical bars on data points depict the statistical uncertainties while the systematic uncertainties are shown as grey boxes. The hadronic cocktail yields from U$+$U collisions are $\sim$5-12\% higher than those from Au$+$Au collisions in given centrality bins, thus only cocktails for Au+Au collisions are shown here as solid lines with shaded bands representing the systematic uncertainties for clarity. (b) The corresponding ratios of data over cocktail.} \label{massSpectra}
%\end{center}
\end{figure}

In Fig.~\ref{massSpectra}(a), the efficiency-corrected $e^{+}e^{-}$ invariant mass spectra in Au$+$Au and U$+$U collisions for pair $p_T<$ 0.15 GeV/$c$ are shown for different centrality bins within the STAR acceptance ($p_T^{e}>$ 0.2 GeV/$c$, $|\eta^{e}|<$ 1, and $|y^{ee}|<$ 1). The corresponding enhancement factors, expressed as ratios of data over hadronic cocktail, are illustrated in Fig.~\ref{massSpectra}(b). The enhancement factors are found to be significant in the most peripheral (60-80\%) collisions, and get less and less so as one goes from peripheral to semi-peripheral (40-60\%) and to semi-central (10-40\%) collisions. Furthermore, the enhancement factors decrease in the low invariant mass region, then rise above $M_{\phi}$ and finally reach maximum around $M_{\rm{J}/\psi}$ for all three centrality bins in both collision systems. The different behaviors in the enhancement factors between low-mass resonances ($\omega$, $\phi$) and J/$\psi$, indicate that the observed excess may be dominated by different processes~\cite{aliceLowPtJpsi, wangmeiTheory}. A dedicated analysis for J/$\psi$ is underway, while this letter focuses on the mass region of 0.4 $<M_{ee}<$ 2.6 GeV/$c^2$.

\begin{figure}[htbp]
%\begin{center}
\includegraphics[keepaspectratio,width=0.5\textwidth]{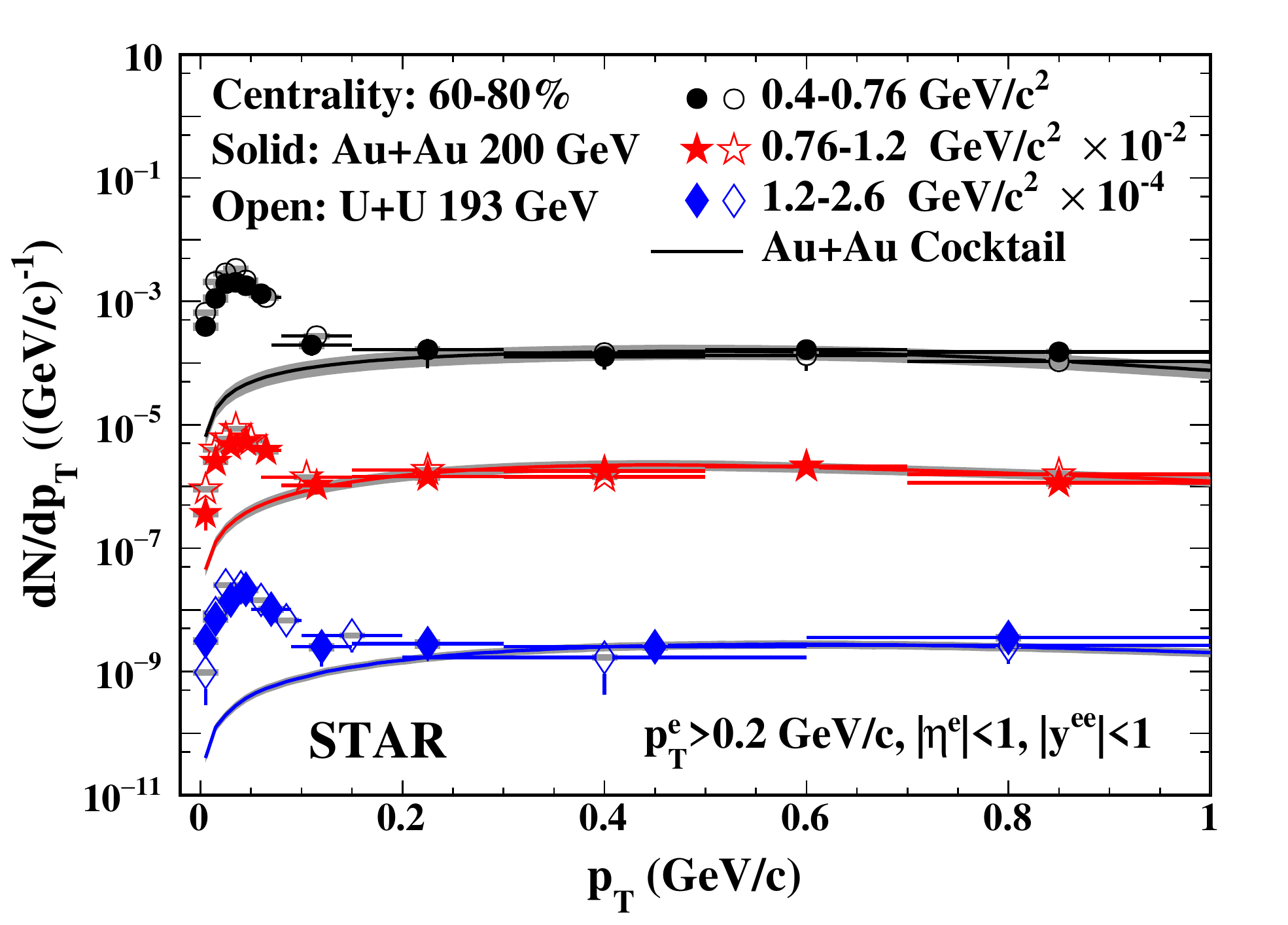}
%\includegraphics[keepaspectratio,angle=270,width=0.5\textwidth]{lowPt_dielectron_ptSpectra.pdf}
%\vspace*{-3mm}
\caption{The $e^{+}e^{-}$ pair $p_T$ distributions within the STAR acceptance for different mass regions in 60-80\% Au+Au and U+U collisions  compared to cocktails. The systematic uncertainties of the data are shown as gray boxes. The gray bands depict the systematic uncertainties of the cocktails.} \label{ptSpectra}
%\end{center}
\end{figure}

\begin{figure}[htbp]
%\centering
\includegraphics[keepaspectratio,width=0.5\textwidth]{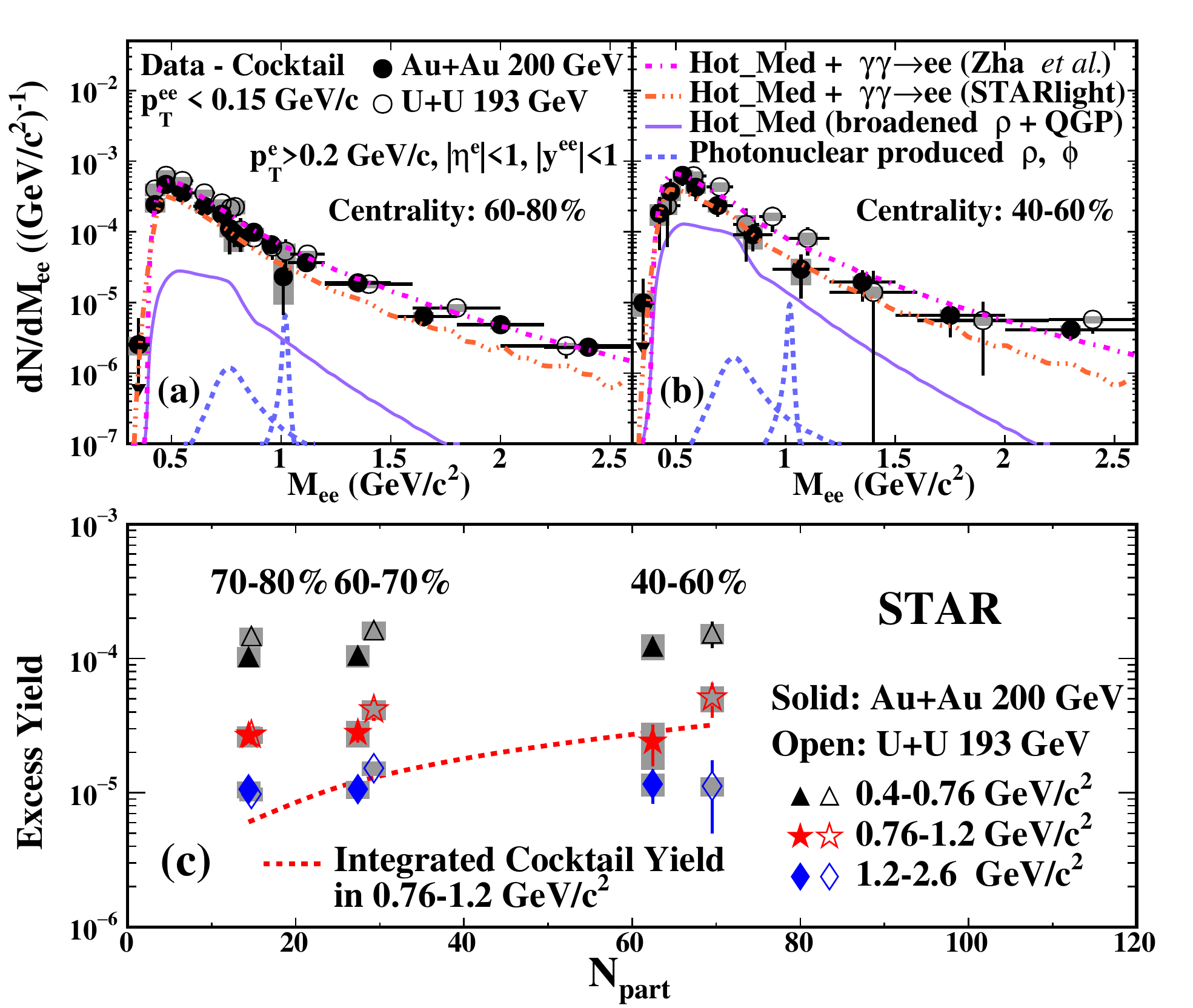}
%\includegraphics[keepaspectratio,angle=270,width=0.5\textwidth]{lowPt_dielectron_excessSpecYield.pdf}
%\vspace*{-3mm}
\caption{The low-$p_T$ ($p_T<$ 0.15 GeV/$c$) $e^{+}e^{-}$ excess mass spectra (data $-$ cocktail) within the STAR acceptance in (a) 60-80\%, (b) 40-60\% for Au+Au and U+U collisions, compared with a broadened $\rho$ model calculation~\cite{broadenRhoTheory}. The contributions of $\rho$, $\phi$ from the photonuclear process are shown, as are the contributions of photon-photon process from two models~\cite{wangmeiTwophoton, starlightTwophoton}. The model calculations are for Au$+$Au collisions in the corresponding centrality bins. (c) The centrality dependence of integrated excess yields in the mass regions of 0.4-0.76, 0.76-1.2, and 1.2-2.6 GeV/$c^{2}$ in Au$+$Au and U$+$U collisions. The centrality dependence of hadronic cocktail yields in the mass region of 0.76-1.2 GeV/$c^{2}$ in both collisions is also shown for comparison. The systematic uncertainties are shown as gray boxes.} \label{excessSpectraYield}
\end{figure}

The $p_T$ distributions of $e^{+}e^{-}$ pairs in three mass regions (0.4-0.76, 0.76-1.2 and 1.2-2.6 GeV/$c^{2}$) are shown in Fig.~\ref{ptSpectra} for 60-80\% Au+Au and U+U collisions, where the enhancement factors are the largest. Interestingly, the observed excess is found to concentrate below $p_T \approx\;$0.15 GeV/c, while the hadronic cocktail, also shown in the figure, can describe the data for $p_T>$ 0.15 GeV/$c$ in all three mass regions.

After statistically subtracting the hadronic cocktail contribution from the inclusive $e^{+}e^{-}$ pairs, the invariant mass distributions for excess pairs for $p_T<$ 0.15 GeV/$c$ are shown in Figs.~\ref{excessSpectraYield}(a) and \ref{excessSpectraYield}(b) for 60-80\% and 40-60\% centralities, respectively. Theoretical calculations incorporating an in-medium broadened $\rho$ spectral function and QGP radiation~\cite{broadenRhoTheory} are also shown in the figures as solid lines. While this broadened $\rho$ model calculation has successfully explained the SPS~\cite{na60Dilepton} and RHIC data~\cite{starDilepton200, starDilepton19.6, phenixDilepton} measured at a higher $p_T$, it cannot describe the enhancement observed at very low $p_T$ in 40-80\% centrality heavy-ion collisions.  We integrated the low-$p_T$ invariant mass distributions for excess pairs over the three aforementioned mass regions and the integrated excess yields are shown in Fig.~\ref{excessSpectraYield}(c) as a function of centrality. Compared to the hadronic cocktail shown as the dashed line in the figure, the excess yields exhibit a much weaker dependence on collision centrality, suggesting that hadronic interactions alone are unlikely to be the source of the excess $e^{+}e^{-}$ pairs.

%\begin{figure*}[htbp]
\begin{figure}[htbp]
\centering
\includegraphics[keepaspectratio, width=0.5\textwidth]{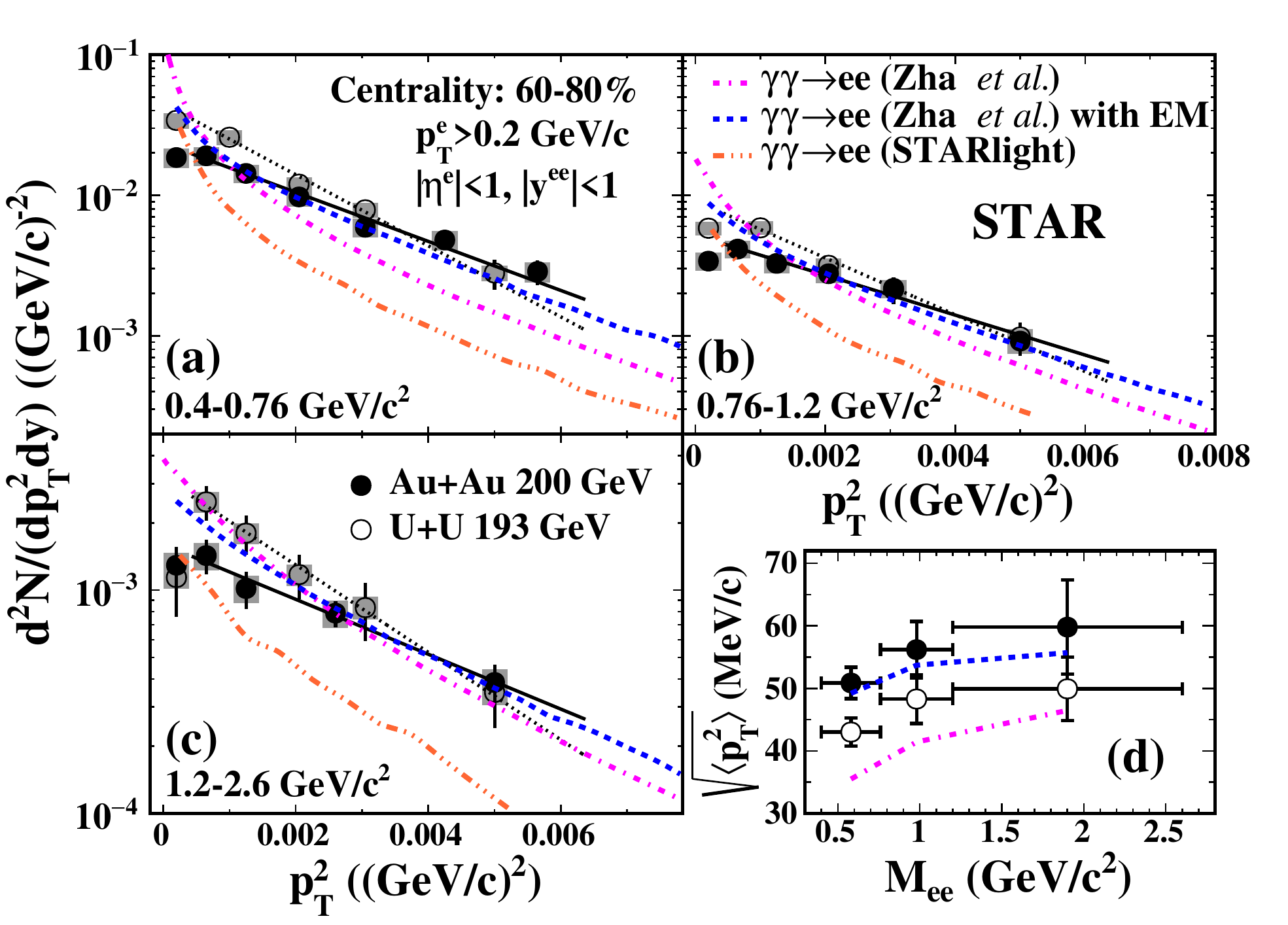}
%\includegraphics[keepaspectratio, angle=270, width=0.5\textwidth]{lowPt_dielectron_tSpectra.pdf}
%\vspace*{-3mm}
\caption{The $p_T^2$ distributions of excess yields within the STAR acceptance in the mass regions of (a) 0.4-0.76, (b) 0.76-1.2, and (c) 1.2-2.6 GeV/$c^{2}$ in 60-80\% Au+Au and U+U collisions. The systematic uncertainties are shown as gray boxes. The solid and dotted lines are exponential fits to the data in Au+Au and U+U collisions, respectively. The dot-dashed and dot-dot dashed lines represent the $p_T^2$ distributions for the photon-photon process from two models~\cite{wangmeiTwophoton, starlightTwophoton} within the STAR acceptance in 60-80\% Au+Au collisions. The dashed lines illustrate the corresponding $p_T^2$ distributions for $e^+e^-$ pairs from model~\cite{wangmeiTwophoton} traversing 1 fm in a constant magnetic field of $10^{14}$ T perpendicular to the beam line. (d) The corresponding $\sqrt{\langle p^2_T\rangle}$ of excess yields. The vertical bars on data points are the combined statistical and systematic uncertainties.} \label{tSpectra}
%\end{figure*}
\end{figure}

In order to investigate the origin of the low-$p_T$ $e^{+}e^{-}$ enhancement, we compared our results to different models~\cite{ wangmeiTheory, wangmeiTwophoton, starlightTwophoton} with the photonuclear and photon-photon contributions employing the equivalent photon approximation (EPA) method~\cite{epaMethod} in Au$+$Au collisions. The model by Zha {\it et al.}~\cite{wangmeiTwophoton} takes into account the charge distribution in the nucleus for estimating the photon flux. Conversely, the model implemented in the STARlight MC generator~\cite{starlightGenerator, starlightTwophoton} treats the nucleus as a point-like charge for evaluating the photon flux and ignores $e^{+}e^{-}$ production within the geometrical radius of the nucleus. Both models assume no effect of hadronic interaction on virtual photon production and do not have uncertainty estimates. The excess based on the model calculations is dominated by photon-photon interactions, in which contributions from Ref.~\cite{wangmeiTwophoton} describe the 60-80\% centrality data fairly well ($\chi^2$/NDF = 19/15, where NDF is the number of degrees of freedom, in 0.4-2.6 GeV/$c^2$), while the results from STARlight underestimate that data ($\chi^2$/NDF = 32/15). In 40-60\% centrality, both models can describe the data within the large statistical uncertainties. The contributions from photonuclear produced $\rho$ and $\phi$ vector mesons, shown as the dashed lines in Figs.~\ref{excessSpectraYield}(a) and~\ref{excessSpectraYield}(b), are found to be negligible. STARlight predicts that the excess yields from photon-photon interactions in U$+$U collisions are $\sim$40\% larger than those in Au$+$Au collisions~\cite{starlightTwophoton}. The observed difference between U+U and Au+Au collisions is consistent with the theoretical prediction within large uncertainties, as shown in Fig.~\ref{excessSpectraYield}(c).

To further explore the low-$p_T$ excess, the $p_T^2$ ($\approx-t$, the squared four-momentum transfer) distributions of the excess yields within the STAR acceptance for 60-80\% centrality are shown in Figs.~\ref{tSpectra}(a)-(c) for three different mass regions. The aforementioned photon-photon model calculations for Au$+$Au collisions are also shown in the figures as dot-dashed and dot-dot dashed lines. The calculations from~\cite{wangmeiTwophoton} fall below data points at large $p_T^2$ values but overshoot data at low $p_T^2$, especially in the extremely low $p_T^2$ region. The calculation from STARlight is lower than that from~\cite{wangmeiTwophoton} but has a similar $p_T$ shape. The spectra dip in data at extremely low $p_T$ ($p_T^2<$ 0.0004 (GeV/$c$)$^2$) and the discrepancy in that $p_T$ region with models could be partially attributed to the EPA method~\cite{epaMethod} without incorporating nonzero photon virtuality~\cite{starUpceepair, qedCalculation}.  Such a discrepancy has been previously observed in the measured low-mass $e^{+}e^{-}$ cross section of photon-photon interactions for $p_T^2<$ 0.000225 (GeV/$c$)$^2$ in UPCs at RHIC~\cite{starUpceepair}. The $\sqrt{\langle p^2_T\rangle}$, which characterizes the $p_T$ broadening, is calculated for both data and aforementioned photon-photon models. In data, a fit of the exponential function ($Ae^{-p_T^2/B^2}$) is performed by excluding the first data points and extrapolated to the unmeasured higher $p_T^2$ region to account for the missing contribution. The uncorrelated systematic uncertainties arising from the raw signal extraction are added in quadrature to the statistical errors, and the resulting total uncertainties are included in the fits. The invariant mass dependence of the extracted $\sqrt{\langle p^2_T\rangle}$ are plotted in Fig.~\ref{tSpectra}(d) for both colliding systems. The $\sqrt{\langle p^2_T\rangle}$ from Au$+$Au collisions are systematically larger than from U$+$U collisions and both increase slightly with increasing pair mass, although the systematic trends are marginally at the level of 1.0-2.3$\sigma$. The values of the $\sqrt{\langle p^2_T\rangle}$ from Au$+$Au data are about 6.1$\sigma$, 3.3$\sigma$, and 1.8$\sigma$ above models~\cite{wangmeiTwophoton, starlightTwophoton} in the 0.4-0.76, 0.76-1.2 and 1.2-2.6 GeV/$c^2$ mass regions, respectively. The general agreements between data and model calculations for $p_{T}$ and invariant mass distributions of $l^+l^-$ pairs produced by photon-photon interactions in UPCs~\cite{starUpceepair, phenixUpceeandJpsi, aliceUpceeandJpsi} are suggestive of possible other origins of the $p_T$ broadening in peripheral collisions as shown in Fig.~\ref{tSpectra}(d). For example, to illustrate the sensitivity the $\sqrt{\langle p^2_T\rangle}$ measurement may have to a postulated magnetic field trapped in a conducting QGP~\cite{cmeConductivity}, we assume each and every pair member generated by model~\cite{wangmeiTwophoton} traverses 1 fm through a constant magnetic field of $10^{14}$ T perpendicular to the beam line ($eBL\approx$ 30 MeV/$c$, where $B$ is $10^{14}$ T, $L$ is 1 fm)~\cite{CMEstrength, Asakawa:2010bu}. The corresponding $p_T^2$ distributions of $e^+e^-$ pairs can qualitatively describe our data except at low $p_T^2$, as shown in Figs.~\ref{tSpectra}(a)-(c). The $\sqrt{\langle p^2_T\rangle}$ of $e^+e^-$ pairs will gain an additional $\sim$30 MeV/$c$, as illustrated in Fig.~\ref{tSpectra}(d). This level of broadening is measurable and may indicate the possible existence of high magnetic fields~\cite{cmeConductivity,CMEstrength,Asakawa:2010bu}.

%% Summary
In summary, we report measurements of $e^{+}e^{-}$ pair production for $p_T<$ 0.15 GeV/$c$ in non-central Au$+$Au collisions at $\sqrt{s_{NN}}$ = 200 GeV and U$+$U collisions at $\sqrt{s_{NN}}$ = 193 GeV. The $e^{+}e^{-}$ yields are significantly enhanced over a wide mass range with respect to the hadronic cocktails in the 40-80\% collisions for both collision species. The entire observed excess is found below $p_T \approx$ 0.15 GeV/c and the excess yield exhibits a much weaker centrality dependence compared to the expectation for hadronic production. The $p^2_T$ distributions of the excess yields in the three mass regions in 60-80\% Au$+$Au and U$+$U collisions are also reported. The $\sqrt{\langle p^2_T\rangle}$ of these distributions show weak invariant mass and collision species dependences. Based on comparisons with model calculations, the observed excess for  $p_T<$ 0.15 GeV/$c$ is very likely linked to photon-photon production and represents the first observation showing the magnitude of two-photon interactions in heavy-ion collisions with hadronic overlap. In addition, model calculations of photon-photon interactions describe the observed excess yields but fail to reproduce the $p_T^2$ distributions.  The level of $p_T$ broadening may indicate the possible existence of a strong magnetic field trapped in a conducting QGP.

We thank the RHIC Operations Group and RCF at BNL, the NERSC Center at LBNL, and the Open Science Grid consortium for providing resources and support. This work was supported in part by the Office of Nuclear Physics within the U.S. DOE Office of Science, the U.S. National Science Foundation, the Ministry of Education and Science of the Russian Federation, National Natural Science Foundation of China, Chinese Academy of Science, the Ministry of Science and Technology of China and the Chinese Ministry of Education, the National Research Foundation of Korea, Czech Science Foundation and Ministry of Education, Youth and Sports of the Czech Republic, Department of Atomic Energy and Department of Science and Technology of the Government of India, the National Science Centre of Poland, the Ministry of Science, Education and Sports of the Republic of Croatia, RosAtom of Russia and German Bundesministerium fur Bildung, Wissenschaft, Forschung and Technologie (BMBF) and the Helmholtz Association.

\bibliography{basename of .bib file}

\end{document}